\documentclass[12pt]{article}
\usepackage{fullpage}
\usepackage[top=2cm, bottom=3cm, left=2cm, right=2cm]{geometry}
\usepackage{amsmath,amsthm,amsfonts,amssymb,amscd}
\usepackage{titlesec}
\usepackage{enumitem}
\usepackage{mathtools}
\usepackage{cancel}
\usepackage{caption}
\usepackage{float}
\usepackage{array}
\usepackage{authblk}

\title{Observations and empirical functions for the ocean surface wave spectrum}
\date{\vspace{-5ex}}

\author[1*]{Hannah Hata Williams}
\author[1]{Michael E. Mueller}
\author[1,2]{Luc Deike}
\affil[1]{Department of Mechanical and Aerospace Engineering, Princeton University, Princeton, New Jersey, USA}
\affil[2]{High Meadows Environmental Institute, Princeton University, Princeton, New Jersey, USA}
\affil[*]{Correspondence: hhwilliams@princeton.edu}

\begin{document}
\maketitle

\abstract{Accurate parameterizations of ocean wave spectra are necessary in a wide array of disciplines including coastal, ocean, and naval engineering as well as in the study of wave interactions and ocean-atmosphere momentum flux. Many such applications use spectrum parameterizations based on temporal data collected well over a half century ago. The development of spatial wave measurement techniques that can accurately capture a larger range of scales allows us to revisit the question of how best to represent an ocean wave spectrum in a variety of ocean wave conditions. We discuss two commonly used wave spectrum parameterizations through a comparison to data collected in field campaigns studying fetch-limited, fully-developed, and mixed sea conditions. We discuss a spectrum parameterization for fully-developed seas that has a $k^{-2.5}$ (or $\omega^{-4}$) dependence on the wavenumber (or angular frequency) in the tail as opposed to the $k^{-3}$ (or $\omega^{-5}$) dependence seen in other frequently-used parameterizations. With knowledge of the peak wavenumber $k_p$ and significant wave height $H_s$, alongside the wind speed, fully-developed conditions can be well-represented. We then compare the impact of using different wave spectrum parameterizations through a Large Eddy Simulation (LES) study of Marine Atmospheric Boundary Layers (MABLs) over the sea surface and find that changing the parameterization used results in variations in the equivalent roughness akin to significant changes in wave conditions.}


\subsection*{Significance Statement}
This study compares three different parameterizations for an ocean wave field (how many waves of each size there are relative to each other given some environmental condition) to updated experimental data taken in a variety of sea states. The most commonly used parameterization for ocean waves is based on theory and data from the 1970s and works well for a specific configuration of wind and waves where the waves are still near the wind source that generated them (fetch-limited). Most of the time---in fully-developed seas---we find that a parameterization based on slightly different theory works better. This is a simple change to make and is relevant to a wide range of applications such as coastal engineering, renewable energy systems (wave energy or offshore wind), wave tank or tow tank design, and more. We demonstrate the implications of the choice of spectrum parameterization in studying the associated marine atmospheric boundary layer flow.

%
\section{Introduction}
Empirical parameterizations for ocean wave energy density spectra are used for a wide variety of ocean and coastal engineering applications as well as for fundamental study of ocean processes, both in experimental and computational contexts. A seminal parameterization for the wave spectrum is the JONSWAP (Joint North Sea Wave Project) spectrum proposed by Hasselmann et al. (1973) \cite{hasselmann_measurements_1973}. Throughout more than 6000 citations, the JONSWAP spectrum has been used in both scientific and engineering contexts. Kumar et al. (2017) used the JONSWAP parameterization to estimate Stokes drift \cite{kumar_bulk_2017}; Liao et al. (2021) did so in simulations of shallow water infragravity waves \cite{liao_analytical_2021}; Craciunescu et al. (2020) used the the parameterization in an experimental study of energy dissipation from breaking waves \cite{craciunescu_wave_2020}; Kimmoun et al. (2021) used it in a study on nonlinear wave group shoaling \cite{kimmoun_experiments_2021}; and Ardhuin et al. (2025) compared long-wave spectra to the JONSWAP parameterization \cite{ardhuin_sizing_2025}. Deskos (2021), Yang et al. (2014), and Aiyer et al. (2024) \cite{Deskos2021, Yang2014, aiyer_dynamic_2024} considered waves described by the JONSWAP spectrum to study or model the impact of ocean waves on momentum flux in the marine atmospheric boundary layer (MABL). In ocean engineering, the JONSWAP spectrum has been used to design floating photovoltaic structures \cite{claus_key_2022}; wave energy technology \cite{guo_review_2021}; and as reference for towing tank experiments in the context of naval engineering 
\cite{stansberg_specialist_2002}. Michalzik et al. (2019) \cite{michalzik_development_2019} and Draycott et al. (2019) \cite{draycott_capture_2019} cite the JONSWAP spectrum in the development of experimental facilities intended for the design and testing of green coastal infrastructure and offshore renewable energy, respectively.

The JONSWAP spectrum was based on a theoretical understanding of wave dynamics developed in the 1950s and 1960s combined with the data collected in a field campaign in the North Sea by Hasselmann et al. (1973) \cite{hasselmann_measurements_1973}. The JONSWAP field experiment focused on a fetch-limited scenario in which the wind came from a single location onshore, allowing the authors to analyze the development of waves as a function of distance as they grew under the single, consistent wind source.

From a more purely theoretical perspective, Phillips (1957) \cite{Phillips1957} proposed a relationship between the omnidirectional wave energy spectrum (integrated over wave angle) and wave frequency, $S(\omega) \propto g^2 \omega^{-5}$, based on a dimensional argument in what he later called the saturation range \cite{phillips_spectral_1985}, where the energy flux should not be related to the wind speed so is solely dependent on the gravitational acceleration $g$ and the wave frequency $\omega$. Pierson and Moskowitz (1964) \cite{pierson_proposed_1964} followed the similarity theory developed by Kitaigorodskii (1962) that also arrived at a $S(\omega) \propto g^2\omega^{-5}$ relationship for fully-developed seas (reference and explanation can be found in Hasselmann et al. (1973) \cite{hasselmann_measurements_1973}) and recognized agreement with the Phillips theory as well. In presenting the JONSWAP study, Hasselmann et al. (1973) \cite{hasselmann_measurements_1973} referred to the same theoretical arguments as Pierson and Moskowitz but extended the parameterization to include dependence on a non-dimensional fetch in a fetch-limited study of developing wave conditions using a peak enhancement factor that concentrates more energy near the peak wavenumber of the spectrum. The Pierson-Moskowitz parameterization takes the wind velocity at 19.5 m above the surface as input; JONSWAP takes the velocity at 10 m and the fetch. Separately, Toba (1973)
\cite{toba_local_1973} proposed a different relationship between the spectral density and wave frequency, $S(\omega) \propto u_* g \omega^{-4}$, based on a similarity argument of the total nondimensional energy and the relationship between energy and significant wave height $H_s$. Phillips (1985) \cite{phillips_spectral_1985} agreed with this new formulation with $S(\omega) \propto \omega^{-4}$ and provided a different theoretical explanation based on the idea that the three terms in the wave energy balance---input from the wind, dissipation by breaking, and transfer from nonlinear resonant wave-wave interactions---will all be of similar magnitude in an equilibrium range and, because this range does not have a characteristic wave frequency, proportional to each other. The high wavenumber asymptote returns to $\omega^{-5}$ dependence in the saturation range; Phillips wrote that this transition $k_n$ occurs at around $4k_p$ (in frequency space, $\omega_n \propto 2\omega_p$) or $k_n \propto g/u_*^2$ while more recent estimates suggest that it scales with the wave age \cite{romero_airborne_2010} or the significant wave height \cite{lenain_measurements_2017}; more discussion on the transition wavenumber can be found in Deike (2022) \cite{deike_mass_2022}. Battjes et al. (1987) \cite{battjes_reanalysis_1987} later reanalyzed the JONSWAP data and found the $\omega^{-4}$ relationship to fit the data better than $\omega^{-5}$. Elfouhaily et al. (1997) \cite{elfouhaily_unified_1997} recognized the varying literature on wave spectrum parameterizations and combined formulae by dividing the wavenumber range into two regimes with $S_{lower}(\omega) \propto \omega^{-4}$ and using factors from both the Pierson-Moskowitz (exponential fall-off) and JONSWAP (peak enhancement) spectra. The high-wavenumber range also has dependence $S_{higher}(\omega) \propto \omega^{-4}$ with a different exponential factor for viscous cutoff from the gravity-capillary limit. This parameterization, while intended for multiple types of sea states, depends on a large number of fitted constants---its complexity makes it more difficult to use so we do not include it in this discussion, for which we prioritize simplicity and ease of implementation with measurable data. 

In summary, the JONSWAP \cite{hasselmann_measurements_1973} and Pierson-Moskowitz \cite{pierson_proposed_1964} wave spectrum parameterizations have a dependence of $S(\omega)\propto \omega^{-5}$ or, equivalently, $\phi(k)\propto k^{-3}$. Both have similar falloffs for long waves, and JONSWAP has an additional peak enhancement factor that concentrates more energy around the peak wavenumber. The spectrum parameterization proposed by Elfouhaily et al. (1997) \cite{elfouhaily_unified_1997} has dependence $S(\omega)\propto\omega^{-4}$ or $\phi(k)\propto k^{-2.5}$, which is in agreement with theory from Toba (1973) \cite{toba_local_1973} and Phillips (1985) \cite{phillips_spectral_1985}. The JONSWAP, Pierson-Moskowitz, and Elfouhaily spectrum parameterizations all use the wind velocity above the surface as an input; JONSWAP and Elfouhaily also use the fetch. A summary of the different parameterizations can be found in Table \ref{tab:spectra}; the JONSWAP and Pierson-Moskowitz parameterizations are discussed in more detail in Section \ref{sec:spectra}.

Many of the field data used in the aforementioned studies to verify the proposed spectrum formulations were from single-point measurement techniques and measured spectra in terms of wave frequency. More recent data acquisition techniques such as LIDAR technology allow for a characterization of the spatial ocean surface wave spectrum, capturing the ocean wave surface over a large spatial extent in addition to increasing small-wave resolution \cite{romero_airborne_2010,lenain_measurements_2017,hogan_observations_2025}. Some of these recent data sets present peak wavelengths of around 100 m and resolve wavelengths down to $\sim$0.4 m, whereas the JONSWAP data resolved frequencies of up to 1 Hz (associated with wavelengths of $\sim$1.5 m).

In this paper we discuss and recommend the use of a parameterization based on the Phillips (1985) \cite{phillips_spectral_1985} theory, herein referred to as the Equilibrium spectrum, for the fully-developed ocean. We compare the performance of three different wave spectrum parameterizations---Pierson-Moskowitz, JONSWAP, and Equilibrium---to LIDAR data collected in a variety of wind-wave conditions. Fetch-limited, or developing, conditions refer to situations in which waves are generated from consistent wind over a finite, measurable distance (the fetch $X$); fetch-limited wave fields usually result in aligned wind and waves. Fully-developed seas refer to conditions where the memory of the fetch and initial storm has been forgotten; these seas are sometimes referred to as having infinite fetch and there is still a main alignment between the wind and wave directions. Mixed wave conditions refer to those that result from a combination of sources coming from multiple directions (and conditions), such as a long distant swell and a local wind-generated sea crossing at an angle, and are prevalent in the presence of obstructions near the coast. We then examine the changes that result from using different "equivalent" wave spectrum parameterizations in a real use scenario through an LES study to demonstrate the sensitivity of momentum transfer and resulting MABL statistics to the choice of spectrum parameterization. 

\begin{table}[t]
    \scriptsize
    \caption{Summary of wave energy spectrum parameterizations comparing applicable sea conditions, dependence on $\omega$ and $k$, input parameters, and constants that have been fitted to data. The parameterization from Elfouhaily et al. (1997) \cite{elfouhaily_unified_1997} is also included here: for this spectrum, $\Omega = U_{10}/c_p$ is the wave age and $F_p$ is the long wave side-effect function for which there is an additional coefficient.}
    \begin{center}
    \begin{tabular}{*{6}{c}}
    \hline
    Parameterization & Conditions & $S(\omega) \propto$ & $\phi(k) \propto$ & Inputs & Constants\\
    \hline
    Pierson-Moskowitz & fully-developed & $\omega^{-5}$ & $k^{-3}$ & $U_{19.5}$ & $\alpha_{PM}$, $\beta_{PM}$\\
    JONSWAP & fetch-limited & $\omega^{-5}$ & $k^{-3}$ & $U_{10}$, $X$ & $\alpha_{JS}$, $\beta_{JS}$, $\gamma$, $\Sigma$, $\omega_p$\\
    Elfouhaily et al. & all & $\omega^{-4}$ for $\omega > 10\omega_p$ & $k^{-2.5}$ for $k < 10k_p$ & $U_{10}$, $X$ & $\alpha_{El}$, $\beta_{PM}$, $\gamma_{JS}$, $\Sigma_{JS}$, $\Omega$, $F_p$\\
    Equilibrium & fully-developed & $\omega^{-4}$ for $\omega > \omega_n$ & $k^{-2.5}$ for $k < k_n$ & $k_p$, $H_s$ & $\beta_E$\\
    \hline
    \end{tabular}
    \end{center}
    \label{tab:spectra}
\end{table}

\section{Wave Spectrum Parameterizations}
\label{sec:spectra}
A discussion of different omnidirectional ocean wave spectrum parameterizations follows. The Pierson-Moskowitz and JONSWAP spectra are presented in the respective literature in terms of angular frequency $\omega$ and can be converted to the wavenumber spectrum $\phi(k)$ by requiring that the total energy in the spectrum is constant through the identity 
\begin{equation}
    \int \phi(k)dk = \int S(\omega) d\omega \text{ .}
\end{equation}
Applying the linear dispersion relation for gravity waves $\omega = \sqrt{gk}$ (and implicitly assuming all energy is contained along the linear dispersion relation) gives
\begin{equation}
    \phi(k) = S(\omega)\frac{d\omega}{dk} = \frac{g}{2\omega} S(\omega) \text{ .}
\end{equation}
The total energy in the spectrum is then related (by definition) to the significant wave height:
\begin{equation} \label{eq:Hs}
  H_s = 4 \sqrt{\int \phi(k)dk} = 4\sqrt{\int S(\omega)d\omega} = 4\sqrt{\langle\eta^2\rangle} \text{ ,}
\end{equation}
where $\eta$ is the instantaneous wave height and the angle brackets denote either a spatial or temporal average (equivalent given a long enough sample period or range). 

\subsection{The Pierson-Moskowitz spectrum}
Pierson and Moskowitz (1964) \cite{pierson_proposed_1964} presented a one-parameter wave spectrum formulation for fully developed seas:
\begin{align}
    S_{PM}(\omega) &= \frac{\alpha_{PM}g^2}{\omega^5}\exp\left[-\beta_{PM}\left(\frac{\omega_0}{\omega}\right)^4\right] \hspace{1em}\text{or} \label{eq:SwPM} \\
    \phi_{PM}(k) &= \frac{\alpha_{PM}}{2k^3}\exp\left[-\beta_{PM}\left(\frac{k_0}{k}\right)^2\right] \text{ ,} \label{eq:EkPM}
\end{align}
where $\omega_0 = g/U_{19.5}$---or, from the dispersion relation, $k_0 = g/U_{19.5}^2$---is dependent on the gravitational constant and the wind velocity measured 19.5 m above the sea surface ($U_{19.5}$ is the single parameter through which this spectrum changes). The constants are fitted to data obtained by Moskowitz (1964) \cite{moskowitz_estimates_1964} and are presented as $\alpha_{PM} = 8.10 \times 10^{-3}$ and $\beta_{PM} = 0.74$. Pierson (1964) \cite{pierson_jr_interpretation_1964} connected this parameterization to the significant wave height through the fitted approximation $H_s \approx 2.14\times10^{-2} U_{19.5}^2$. The low wavenumber dependence on $k$ is governed by the exponential factor, and the high wavenumber tail is governed by the $k^{-3}$ dependence.

\subsection{The JONSWAP spectrum}
Hasselmann et al. (1973) \cite{hasselmann_measurements_1973} proposed an updated formulation based on the fetch-limited study using data obtained in the Joint North Sea Wave Project (JONSWAP):
\begin{align}
    S_{JS}(\omega) &= \frac{\alpha_{JS}g^2}{\omega^5}\exp\left[-\beta_{JS}\left(\frac{\omega_p}{\omega}\right)^4\right] \gamma^{\exp\left[\frac{\left(\omega-\omega_p\right)^2}{2\Sigma^2\omega_p^2}\right]} \hspace{1em}\text{or} \label{eq:SwJS}\\
    \phi_{JS}(k) &= \frac{\alpha_{JS}}{2k^3}\exp\left[-\beta_{JS}\left(\frac{k_p}{k}\right)^2\right] \gamma^{\exp\left[\frac{\left(\sqrt{k}-\sqrt{k_p}\right)^2}{2\Sigma^2k_p}\right]} \text{ ,} \label{eq:EkJS}
\end{align}
where $\beta_{JS} = \frac{5}{4}$. This formulation, in contrast to the Pierson-Moskowitz one, is dependent on two environmental parameters: the velocity 10 m above the sea surface $U_{10}$ and the fetch $X$. The other parameters are tied to both $U_{10}$ and $X$ through a three-way coupling:
\begin{equation}
    \label{eq:js_omega}
    \omega_p = 22\left(\frac{g^2}{U_{10}X}\right)^{1/3} \text{ , which converts to } 
\end{equation}
\begin{equation}
    k_p = 484g^{-1/3}(U_{10}X)^{-2} \text{ , and}
\end{equation}
\begin{equation}
    \alpha_{JS} = 0.076\left(\frac{U_{10}^2}{Xg}\right)^{0.22} \text{ .}
\end{equation}
The authors also included a relationship between the two input parameters and the total energy in the spectrum, 
\begin{equation}
    \label{eq:js_int}
    \int S(\omega)d\omega = 1.6\times10^{-7} \frac{U_{10}^2 X}{g} \text{ .}
\end{equation} 
The JONSWAP formulation is similar to Pierson-Moskowitz in the shape of the tail (high wavenumber dependence on $k^{-3}$) and the falloff (low wavenumber dependence governed by the exponential factor) with the addition of a peak enhancement factor $\gamma$ = 3.3, with $\Sigma = 0.07$ if $k \leq k_p$ and $\Sigma = 0.09$ if $k > k_p$. These constants were fit to the data collected as part of the JONSWAP experiment.

\begin{table}[t]
    \scriptsize
    \caption{Wind and wave conditions from data taken in fully-developed seas during the SoCal 2013 \cite{lenain_measurements_2017}, HiRes2010 \cite{sutherland_field_2013}, and ASIT TKE \cite{hogan_observations_2025} experimental campaigns. The reported friction velocity is used when available; all other parameters are calculated from the wave energy spectrum. The peak wavelength $\lambda_p$ is included for reference.}
    \begin{center}
    \begin{tabular}{c | c | *{6}{c}}
    \hline\hline
    & \multicolumn{1}{c|}{Reported} & \multicolumn{6}{c}{Calculated}\\
    Dataset & $u_*$ & $k_p$ & $\lambda_p$ & $H_s$ & $k_n$ & $c_p/u_*$ & $H_sk_p$ \\
    & m/s & rad/m & m & m & rad/m & - & - \\
    \hline
    SoCal2013 & - & 0.063 & 100 & 2.40 & 0.42 & - & 0.15 \\
    HiRes 2010, 1 & 0.535 & 0.031 & 203 & 4.10 & 0.77 & 33.3 & 0.13\\
    ASIT TKE 164 & 0.576 & 0.054 & 116 & 3.55 & 0.50 & 23.4 & 0.19\\
    ASIT TKE 50 & 0.251 & 0.072 & 87 & 1.46 & 1.20 & 46.5 & 0.10\\ 
    \hline
    \end{tabular}
    \end{center}
    \label{tab:wave_params_fullyd}
\end{table}

\begin{table}[t]
    \scriptsize
    \caption{Wind and wave conditions from the GOTEX \cite{romero_airborne_2010} field campaign in fetch-limited seas. Values of $U_{10}$ and $X$ calculated to find the best fit with the Pierson-Moskowitz and JONSWAP spectra are shown in comparison to reported conditions ($U_{19.5}$ has been converted to $U_{10,PM}$ for the Pierson-Moskowitz spectrum using the log law and the reported friction velocity). }
    \begin{center}
    \begin{tabular}{c | *{3}{c} | *{8}{c}}
    \hline\hline
    & \multicolumn{3}{c|}{Reported} & \multicolumn{8}{c}{Calculated}\\
    Dataset & $U_{10}$ & $X$ & $u_*$ & $U_{10,PM}$ & $U_{10,JS}$ & $X$ & $k_p$ & $\lambda_p$ & $H_s$ & $c_p/u_*$ & $H_sk_p$ \\
    & m/s & m & m/s & m/s & m/s & km & rad/m & m & m & - & - \\
    \hline
    GOTEX RF05\_1 & 18.14 & 7 & 0.747 & 5.90 & 16.89 & 16 & 0.250 & 25 & 1.08 & 8.40 & 0.27 \\
    GOTEX RF05\_2 & 18.13 & 12 & 0.747 & 7.22 & 19.83 & 23 & 0.177 & 35 & 1.52 & 10.0 & 0.27 \\
    GOTEX RF05\_4 & 18.09 & 24 & 0.748 & 8.63 & 18.77 & 47 & 0.113 & 56 & 2.08 & 12.5 & 0.23 \\
    GOTEX RF05\_8 & 17.97 & 63 & 0.753 & 10.03 & 18.60 & 81 & 0.079 & 80 & 2.71 & 14.8 & 0.21\\
    GOTEX RF05\_12 & 15.16 & 354 & 0.681 & 12.42 & 13.70 & 313 & 0.039 & 161 & 3.92 & 23.3 & 0.15\\
    \hline
    \end{tabular}
    \end{center}
    \label{tab:wave_params_fetchl}
\end{table}

\subsection{The Equilibrium Spectrum}
The Equilibrium wave spectrum formulation is based on the concept of equilibrium between the wind input into waves, nonlinear wave interactions, and energy dissipation by breaking discussed by Toba (1973) \cite{toba_local_1973} and by Phillips (1985) \cite{phillips_spectral_1985}. The energy in the equilibrium range is described to vary with $u_* g^{-1/2} k^{-5/2}$, which can be combined with an exponential falloff factor for low wavenumbers \cite{elfouhaily_unified_1997,wu_breaking_2023,wu_turbulence_2025}:
\begin{equation}\label{eq:EkE}
    \phi_{E}(k) = P u_* g^{-1/2} k^{-5/2}\exp\left[-\frac{5}{4}\left(\frac{k_p}{k}\right)^2\right] \text{ .}
\end{equation}
The prefactor $P$ is constant in both Toba's and Phillips' parameterizations but can vary with wave age \cite{romero_airborne_2010} and therefore fetch \cite{elfouhaily_unified_1997}. The equilibrium spectrum is compatible with the description from weak-wave turbulence theory based on nonlinear wave interactions driving a forward wave energy cascade and leading to the relation $\phi(k)\propto \mathcal{P}^{1/3} g^{-1/2} k^{-5/2}$, where $\mathcal{P}$ is the nonlinear wave energy flux and has dimensions of velocity (wind speed) cubed \cite{zakharov_kolmogorov_1992}.

In the following, we set the prefactor $P$ based on the desired significant wave height $H_s$ through its relationship to the total energy in the spectrum according to equation \ref{eq:Hs}. While $u_\ast$ is included in the full parameterization in equation \ref{eq:EkE} to describe the wind forcing, knowledge of the total wave energy through $H_s$ is used as a diagnostic to determine the prefactor $P$. Although the number of input parameters used to describe the spectrum is the same as the JONSWAP parameterization, the inputs control distinct and independent characteristics of the wave energy spectrum, whereas with JONSWAP the fetch $X$ and wind speed $U_{10}$ both control the peak frequency and significant wave height through the (coupled) fetch-limited relationships.

The Equilibrium spectrum is expected to be valid for $k<k_n$, a transition wavenumber beyond which the waves are in the saturation range and are governed by $\phi_{E}(k) \propto k^{-3}$. We use the low-wavelength exponential factor originally proposed by Pierson and Moskowitz (1964) \cite{pierson_proposed_1964} because it represents data presented here (with resolution at low wavenumbers down to typically 0.03 rad/m) relatively well. A formulation for the large wave fall-off has recently been proposed by Ardhuin et al. (2025) \cite{ardhuin_sizing_2025} with a sharper power law fitted to data with high resolution of long wavelengths (greater than 500m i.e., $k$ less than 0.01 rad/m) not accessible in the LIDAR data discussed here. The main focus of this discussion is on the high-$k$ part of the spectrum.

\begin{figure}
    \begin{center}
    \noindent\includegraphics[width=0.8\textwidth]{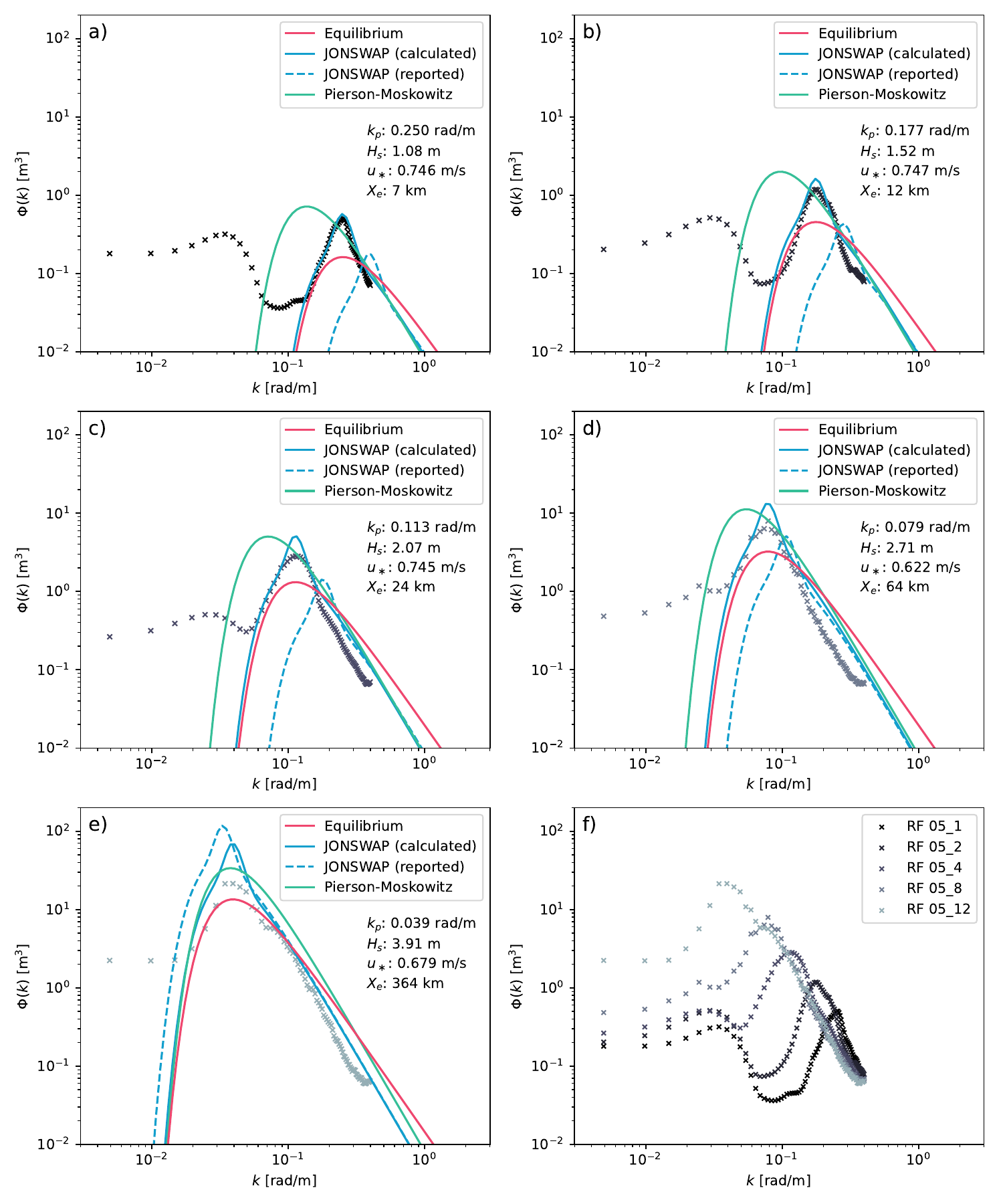}\\
    \caption{Omnidirectional wave spectra computed using JONSWAP, Pierson-Moskowitz, and Equilibrium (solid lines) compared to LIDAR data obtained in the fetch-limited GOTEX campaign (marked by 'x') for a range of fetches from Romero and Melville (2010) \cite{romero_airborne_2010}. The JONSWAP wave spectrum according to the reported values of $U_{10}$ and $X$ is plotted as a dashed line for comparison. The conditions span peak wavenumbers of 0.039-0.250 rad/m, corresponding to peak wavelengths of 25-161 m. The JONSWAP parameterization captures the shape of the lowest-fetch case very well when intentionally matched to the conditions but reported values do are not able to capture the location or magnitude of the peak. The peak shape also decreases in accuracy as the fetch increases (a-e). The bottom right panel (f) shows the GOTEX data for all peak wavenumbers together.}
    \label{fig:fetch}
    \end{center}
\end{figure}

\section{Wave Spectrum Parameterization Performance Compared to recent field data}
\label{sec:results}
In this section we compare JONSWAP, Pierson-Moskowitz, and Equilibrium wave spectrum parameterizations to data collected in four field campaigns. We consider a representative selection of recent high quality spatial measurements of surface wave spectra capturing a wide range of wave scales. 

The GOTEX experiment, presented by Romero and Melville (2010) \cite{romero_airborne_2010}, is a fetch-limited study performed in the Gulf of Tehuantepec with a conically-scanning LIDAR mounted on aircraft reporting wavenumbers up to around 0.5 rad/m ($\approx$ 12.5 m wavelength). These measurements are taken in winds with $U_{10}$ = 15-18 m/s and include wind-waves with significant wave heights from 1.1 to 3.9 m ($H_s$ calculated from reported values of $\langle\eta^2\rangle$ and extracted from energy spectra were similar).

The HiRes 2010 experiment---discussed in Sutherland and Melville (2013) \cite{sutherland_field_2013}, Sutherland and Melville (2015) \cite{sutherland_field_2015}, and Pizzo et al. (2019) \cite{pizzo_lagrangian_2019}---uses a research platform with instrumentation including infrared cameras and waveform scanning LIDAR resolving wavenumbers up to around 10 rad/m (i.e. wavelengths down to $\approx$ 0.5 m). The SoCal2013 experiment, presented by Lenain and Melville (2017) \cite{lenain_measurements_2017}, uses the same research platform as in the HiRes 2010 experiment in addition to waveform scanning LIDAR mounted on aircraft; wavenumber resolution given by these data is similarly up to around 10 rad/m. Both the HiRes 2010 and SoCal2013 campaigns were performed in fully-developed seas in the Pacific Ocean off of Northern and Southern California, respectively. 

The ASIT TKE experiment, presented by Hogan et al. (2025) \cite{hogan_observations_2025}, uses the Woods Hole Air-Sea Interaction tower equipped with laser altimeters, visible cameras, an infrared camera, and a polarimetric imager to capture the ocean surface. The data were taken on the Atlantic Shelf, approximately 3 km off the coast of Martha's Vineyard, and resolve wavenumbers up to around 6 rad/m ($\approx$ 1 m wavelength).

These three relatively recent data sets permit a discussion regarding a broad range of wave energy spectra in fully developed states, fetch limited states, and complex coastal environments.

To compare field data to the wave spectrum parameterizations described in Section \ref{sec:spectra}, we extract the peak wavenumber $k_p$ from the maximum in the given dataset and calculate the significant wave height $H_s$ based on the total energy (according to equation \ref{eq:Hs}). Calculated wave field parameters for fully-developed sea and mixed environment cases can be seen in Table \ref{tab:wave_params_fullyd}; parameters for fetch-limited cases are in Table \ref{tab:wave_params_fetchl}. We determine the Pierson-Moskowitz and JONSWAP input parameters $U_{19.5}$, $U_{10}$, and $X$ through the relationships to the peak wavenumber $k_p$ and significant wave height $H_s$ (or, equivalently, the total energy) presented by each respective source and described in Section \ref{sec:spectra}. $U_{10}$ and $X$, when reported, are also compared to the input values calculated in Table \ref{tab:wave_params_fetchl} (we calculate an equivalent $U_{10}$ for the Pierson-Moskowitz spectrum using the log law for this comparison). Extracting the parameters based on the peak wavenumber and total wave energy in this way gives a best case scenario for these spectrum parameterizations: comparisons with inputs from reported conditions (such as $U_{10}$) resulted in wave spectra that differed more from the data. Furthermore, calculating the equivalent spectra in this way allows for standardization across all datasets since those collected in fully-developed seas do not measure a fetch. The Equilibrium spectrum uses the two extracted parameters $k_p$ and $H_s$ directly, and we determine the transition wavenumber $k_n$ for the Equilibrium spectrum empirically. $u_\ast$ is also part of the Equilibrium spectrum parameterization but since we adjust the prefactor $P$ based on the significant wave height, its information is redundant. In this way, we can use the parameterization as a diagnostic for the shape of the spectrum.

\subsection{Fetch-limited conditions}

Figure \ref{fig:fetch} shows wave spectra in fetch-limited wave conditions from Romero and Melville (2010) \cite{romero_airborne_2010} for a range of equivalent fetches from 7 to 364 km. The JONSWAP parameterization---which was developed for a fetch-limited case---reproduces the energy spectrum very well for the smallest fetch cases when specifically fit to calculated $k_p$ and $H_s$, highlighting the importance of the peak enhancement idea in fetch limited conditions. For these small-fetch cases, the Equilibrium parameterization is unable to capture the sharp peak, and the Pierson-Moskowitz parameterization significantly misestimates where the energy peak is. As the fetch increases (and the conditions grow closer to that of a fully-developed sea state), JONSWAP overestimates the peak, and the other two parameterizations become more accurate, although Pierson-Moskowitz continues to overestimate the total energy in the spectrum due to having only a single input parameter $U_{19.5}$. The JONSWAP parameterization using reported values of $U_{10}$ and $X$ consistently misses both the wavenumber and magnitude of the peak, especially for low-fetch cases where the shape is most accurate.

\begin{figure}[t]
    \begin{center}
    \noindent\includegraphics[width=0.48\textwidth]{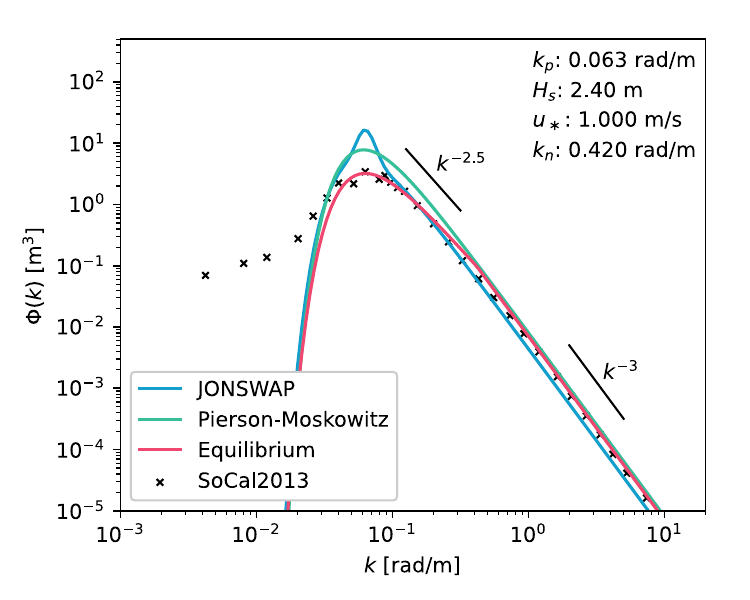}
    \noindent\includegraphics[width=0.48\textwidth]{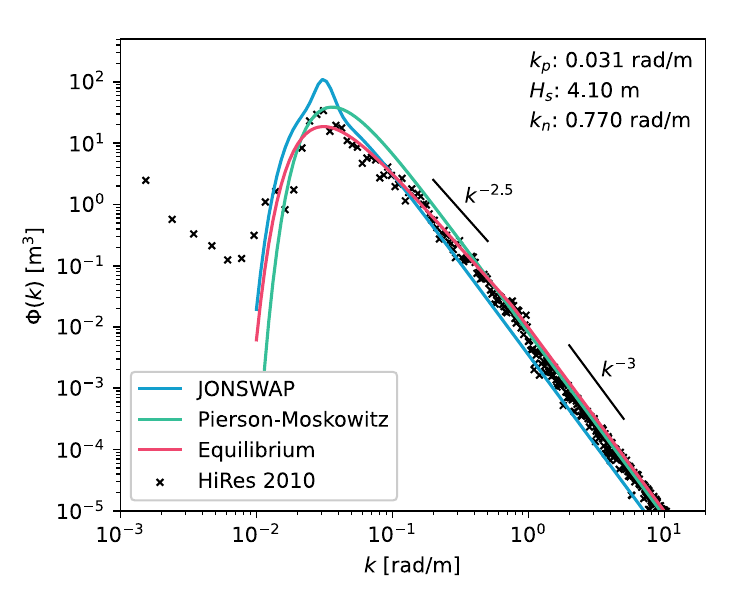}
    \caption{Omnidirectional wave spectra for fully developed wave fields obtained by LIDAR in SoCal2013 (left) \cite{lenain_measurements_2017} and HiRes 2010 (right) \cite{sutherland_field_2013} compared to JONSWAP, Pierson-Moskowitz, and Equilibrium parameterizations (solid lines). For these fully-developed sea conditions, the peak enhancement factor in the JONSWAP parameterization does not accurately capture the shape of the profile. The case from HiRes 2010 has reported values of $U_{10}$ = 12.9 m/s, $u_*$ = 0.535 m/s, and $H_s$ = 4.15 m (similar to extracted estimate of 4.10 m). Values of $k_p$ and $H_s$ for the SoCal 2013 data are extracted from the profile; $U_{10}$ and $u_*$ are unknown. The transition wavenumbers for the Equilibrium parameterization is determined empirically.}
    \label{fig:fdseas}
    \end{center}
\end{figure}

\subsection{Fully-developed conditions}

Figure \ref{fig:fdseas} shows samples of the SoCal2013 and HiRes 2010 campaigns for fully-developed sea conditions. For these conditions, the Equilibrium wave energy spectrum captures the shape of the field measurements near the peak and through the high-wavenumber tail, where there is initially a $k^{-5/2}$ dependence that transitions into a $k^{-3}$ one. The Pierson-Moskowitz parameterization's peak is more aligned in these fully-developed sea cases than in the fetch-limited scenario but still overestimates the total energy. The JONSWAP parameterization's peak enhancement appears to be over-emphasized in these conditions, and the tail falls off faster than the data; these traits have the combined impact of concentrating more energy near the peak than is observed.

\begin{figure}[t]
    \begin{center}
    \noindent\includegraphics[width=0.95\textwidth]{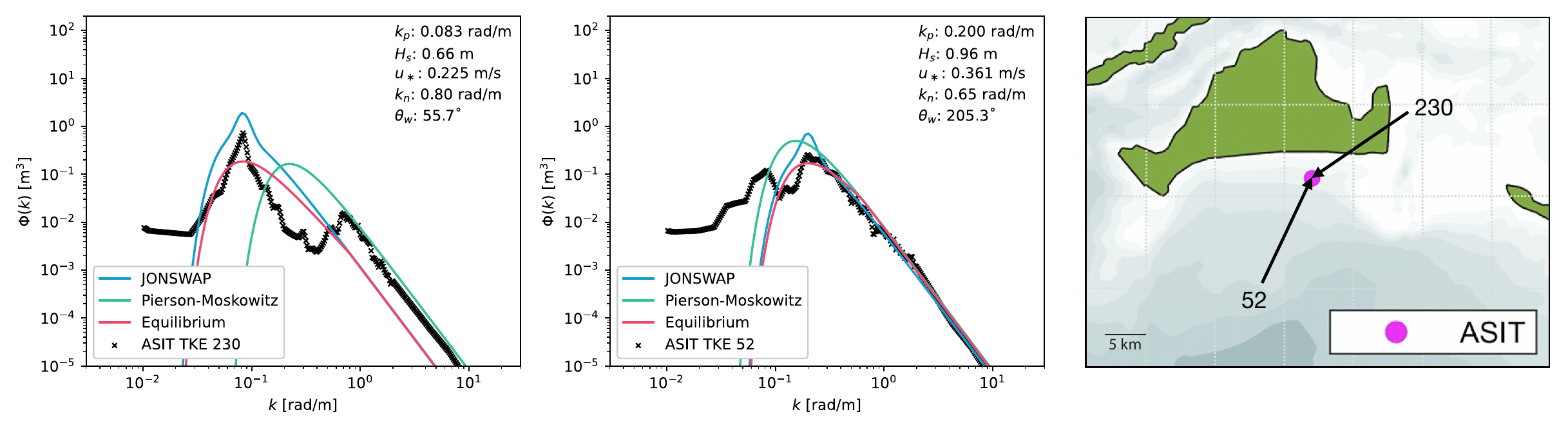}
    \caption{Omnidirectional wave spectra computed using JONSWAP, Pierson-Moskowitz, and Equilibrium parameterizations (solid lines) compared to data obtained for two cases from the ASIT TKE experiment (marked by 'x') collected in 2019 and presented by Hogan et al. (2025) \cite{hogan_observations_2025} along with map showing the wind direction for each of the two cases (right). The left plot shows an offshore case (wind comes off the island) case with a double-peak indicating a complex sea state. The center plot shows an onshore wind case (wind comes off the ocean) that is closer to a fully-developed sea state.}
    \label{fig:hogan}
    \end{center}
\end{figure}

\subsection{Mixed sea state}

Figure \ref{fig:hogan} compares wave spectrum parameterizations to data collected as part of the ASIT TKE experiment \cite{hogan_observations_2025}. Since the experiment was conducted very close to the shore of an island, there is a significant impact of wind direction on the resulting wave field. Wind coming from angles between 270°-90° (measured from North = 0°) are offshore winds, meaning that they blow from the direction of the island. The left panel of Figure \ref{fig:hogan} is one such case, where a double-peak in the wave spectrum indicates a combined wave field with some wind-waves developing from the prevailing wind on top of existing waves generated by other winds. The tower used in the ASIT TKE experiment was at a water depth of 19 m, which makes the deep water linear dispersion relation less accurate, further impacting similarity of the spectra from the experiment to deep water parameterizations. None of the three wave spectrum parameterizations are able to capture this double peak since they were not designed to account for a mixed sea state in which the waves come from multiple directions. The slope at high-wavenumber does appear to be consistent with the expected $k^{-3}$ relationship. The center panel shows the wave spectrum from an onshore wind that blows from the direction of the ocean and has a more familiar shape. While omnidirectional spectra can represent some wave conditions very close to the shore, these data demonstrate that others are more complicated and cannot be accounted for by the parameterizations presented here.

\section{Impact of Spectrum in a Large Eddy Simulation of Marine Atmospheric Boundary Layer}
\label{sec:les}

\begin{figure}[t]
    \begin{center}
    \noindent\includegraphics[width=0.8\textwidth]{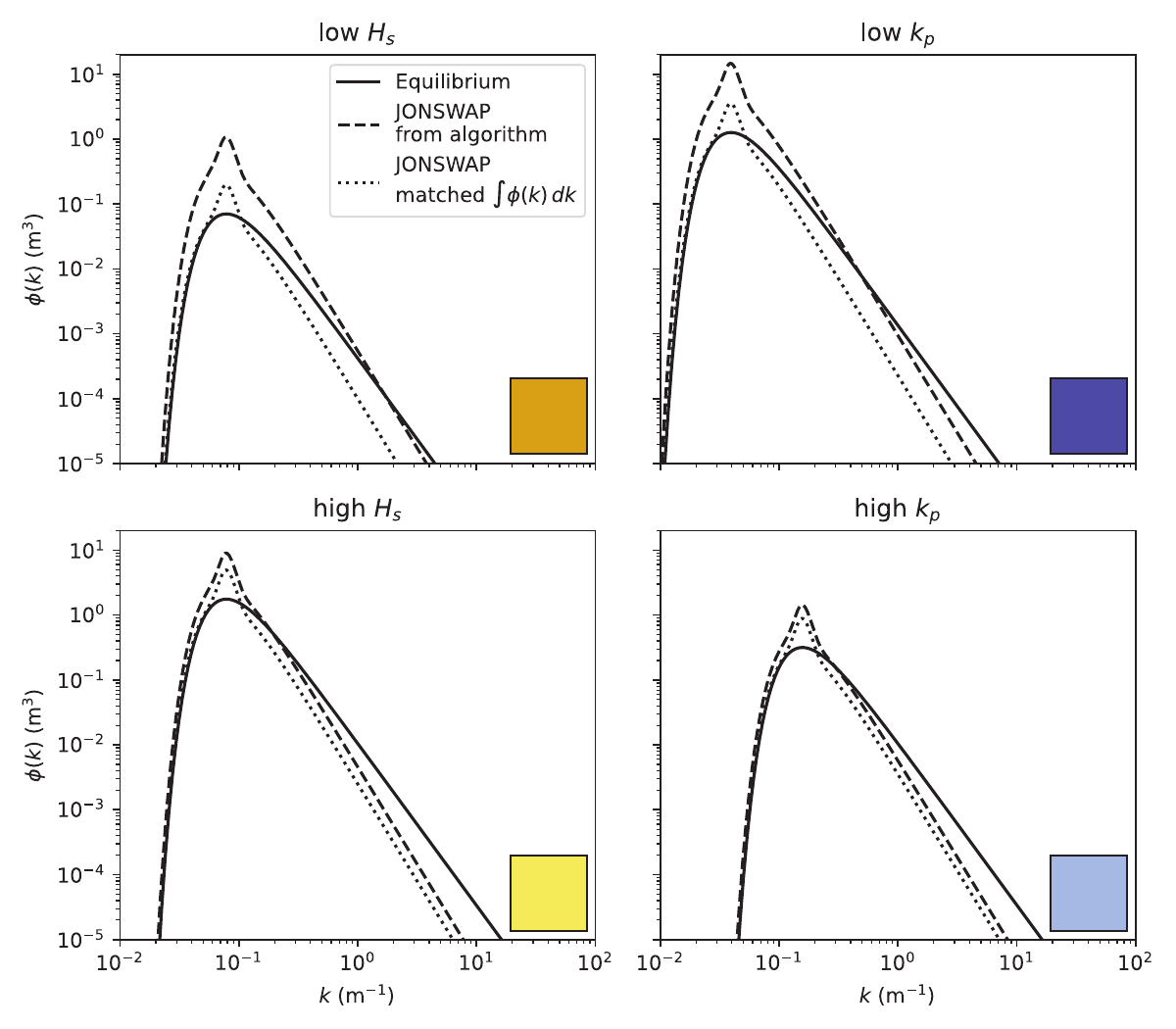}
    \caption{Spectra for wave conditions used in LES comparison study, described in Table \ref{tab:les_cases}. Color coding shown in the lower right hand of each panel will be used in later figures.}
    \label{fig:les_spectra}
    \end{center}
\end{figure}

We conducted an LES study to assess whether changes in spectrum parameterization make a substantial difference on the MABL for wave conditions that are nominally the "same". Four wave conditions spanning ranges of commonly-observed sea states were selected and are described in Table \ref{tab:les_cases}. For each wave condition, the Equilibrium spectrum was generated directly from $H_s$ and $k_p$. Given the input parameters of the JONSWAP formulation, there is no strict equivalent to the Equilibrium spectrum conditions. We consider two JONSWAP spectra to assess the sensitivity of the MABL to this decision. One uses the formulae presented in Hasselmann et al. (1973) \cite{hasselmann_measurements_1973} to convert $H_s$ and $k_p$ into input values $U_{10}$ and $X$, as is done in Section \ref{sec:results} (equations \ref{eq:js_omega}-\ref{eq:js_int}). The second involved doing the same for $k_p$ and iteratively calculating $X$ and $U_{10}$ until the total energy in the spectrum (and therefore $H_s$) matched the target parameters. Whether deriving an equivalent spectrum through the presented algorithm or matching total energy results in a spectrum more similar to the Equilibrium depends on the wave condition, as illustrated by Figure \ref{fig:les_spectra}. We elected to focus on the comparison between Equilibrium and JONSWAP spectrum parameterizations under MABLs because the JONSWAP spectrum is used frequently for both engineering and scientific analysis and because the conclusions on the sensitivity of the MABL to spectrum choice would be similar. We insist that the goal is not to argue which specific wave spectrum parameterization should be used but to investigate the parameterization's impact on practical implications within modeling frameworks such as the MABL profile obtained from LES. The LES and wave model framework is presented in details in a separate paper \cite{williams_sea_2026}; we introduce here a summary of the methodology.

\subsection{LES Framework and Simulation Setup}

\begin{table}[t]
    \scriptsize
    \caption{Wave parameters for the four spectra used to assess impact of spectrum parameterization on MABL development using LES. JONSWAP spectrum input parameters calculated from the formulae (algorithm) provided by Hasselmann et al. (1973) \cite{hasselmann_measurements_1973} are labeled with subscript $a$ while those calculated by matching the total energy are labeled with subscript $m$. The nondimensional parameters $c_p/u_\ast$ and $c_p/U_{10}$ are calculated from the simulation using the Equilibrium spectrum and bulk velocity $U_{bulk} = 18$ m/s. We ran simulations with each of the four wave conditions at three bulk wind velocities ($U_{bulk} = 12, \:18,$ and $24$ m/s) and three parameterizations for a total of 24 cases.}
    \begin{center}
    \begin{tabular}{c | *{4}{c} | *{4}{c} | *{3}{c} }
    \hline\hline
    & \multicolumn{4}{c|}{Input wave parameters} & \multicolumn{4}{c|}{JONSWAP input parameters} & \multicolumn{3}{c}{Nondimensional values (Eq.)} \\
    \hline
    Label & $H_s$ & $k_p$ & $\lambda_p$ & $c_p$ & $X_a$ & $U_{10,a}$ & $X_m$ & $U_{10,m}$ & $H_sk_p$ & $c_p/u_{\ast}$ & $c_p/U_{10}$ \\
    & m & m$^{-1}$ & m & m/s & km & m/s & km & m/s & - & - & - \\
    \hline
    low $H_s$ & 0.4 & 0.0785 & 80 & 11.18 & 3,750 & 0.40 & 49,400 & 0.03 & 0.031 & 24.09 & 0.83 \\
    low $k_p$ & 1.2 & 0.0393 & 160 & 15.80 & 3,320 & 1.29 & 28,600 & 0.15 & 0.047 & 26.93 & 1.29 \\
    high $H_s$ & 2.0 & 0.0785 & 80 & 11.18 & 150 & 10.11 & 375 & 4.04 & 0.157 & 14.96 & 1.10 \\
    high $k_p$ & 1.2 & 0.157 & 40 & 7.91 &  52 & 10.29 & 108 & 4.98 & 0.188 & 10.72 & 0.78 
    \\
    \hline
    \end{tabular}
    \end{center}
    \label{tab:les_cases}
\end{table}

LES calculations were performed using NGA, a structured, finite difference, low Mach number flow solver using second-order centered schemes for spatial discretization and a second-order, iteratively-implicit scheme for temporal integration \cite{Desjardins2008, MacArt2016}. The air velocity field is described by the filtered Navier-Stokes equations in the incompressible limit,
\begin{equation*}
     \nabla \cdot \widetilde{\mathbf{u}} = 0 \text{ ,}\hspace{2em}\text{and} \hspace{3em}
    \frac{\partial\widetilde{\mathbf{u}}}{\partial t} + \widetilde{\mathbf{u}} \cdot \nabla\widetilde{\mathbf{u}} = - \frac{1}{\rho} \nabla \widetilde{p} + \nabla\cdot\widetilde{\underline{\underline{\boldsymbol{\sigma}}}} + \widetilde{\mathbf{F}}_d \text{ ,}
\end{equation*}
which are discretized on a Cartesian grid ($x$,$y$,$z$) where $x$ is the streamwise, $y$ is the spanwise, and $z$ is the vertical direction. Here $\widetilde{\mathbf{u}} = (\widetilde{u},\widetilde{v},\widetilde{w})$ is the velocity vector, where the tilde denotes variables filtered on the LES grid. $\widetilde{\sigma}_{ij} = 2\nu \widetilde{S}_{ij} + \widetilde{\tau}_{ij}^d$ is the total deviatoric stress, where $\nu$ is the molecular viscosity, $\widetilde{S}_{ij}$ is the resolved strain rate tensor, and $\widetilde{\tau}_{ij}^d$ is the subfilter stress (SFS) tensor. The SFS tensor is modeled using a Lilly-Smagorinsky type subfilter viscosity model $\widetilde{\tau}_{ij}^d = 2\nu_T \widetilde{S}_{ij}$, where the subfilter viscosity is computed using the Anisotropic Minimum Dissipation (AMD) model \cite{abkar_large-eddy_2017, rozema_minimum-dissipation_2015}.

Wave drag is calculated using the Wave Spectrum Drag Model \cite{aiyer_sea_2023, aiyer_dynamic_2024, williams_sea_2026}. $\widetilde{\mathbf{F}}_d$ is the resolved wave drag, or the drag force representing the effects of waves larger than the LES grid size, and is applied in the streamwise direction. This drag force is imparted on each grid cell at the bottom of the domain is computed as a sum of the drag contributions of each wave mode so $\widetilde{\mathbf{F}}_d = \sum_{k_n = k_{min}}^{2\pi/\Delta_x} F_{d,n}(x,y,t,k_n)$ where
\begin{equation}
    \label{eq:res_waves}
    F_{d,n} = F_{d}(x, y, t, k_n) = \begin{cases}
    C_{d,n} \frac{1}{\Delta_z} \widetilde{u}_b (\widetilde{u}_b - c_n) \frac{\partial\widetilde{\eta}_n}{\partial x} \mathcal{H}\left\{\frac{\partial\widetilde{\eta}_n}{\partial x}\right\} & \text{if } \widetilde{u}_b > c_n \\
    \beta(k_n)\frac{(a_nk_nu_\ast)^2}{2} & \text{if } \widetilde{u}_b < c_n \text{, } \frac{c_n}{u_\ast} > 25 \text{, and } k_n < 2k_p \\
    0 & \text{otherwise.}
    \end{cases}
\end{equation}
In equation \ref{eq:res_waves}, $\Delta_z$ is the local vertical grid size, $\widetilde{u}_b = \widetilde{u}_b(x, y, t)$ is the filtered streamwise velocity at the first (bottom) grid point above the sea surface, and $c_n = \sqrt{g/k_n}$ is the wave mode phase velocity. The Heaviside function $\mathcal{H}[x]$ ensures that the force is only applied when the flow is incident on the wave frontal area. The drag coefficient $C_{d,n} = a_nk_n/(1+6(a_nk_n)^2)$ is dependent on the wave mode steepness $a_nk_n$. The wave input rate for swell waves (those with $c_n/u_\ast > 25$) is parameterized through an empirical fit to phase-resolved simulations: $\beta(k_n) = 25 - \frac{c_n}{u_\ast}$  \cite{cao_numerical_2021}. The amplitude for each wave mode is defined by the energy spectrum, where $a_n = \sqrt{2\phi(k_n)dk_n}$ and the mode's filtered wave height is $\widetilde{\eta}(x,t) = \sum_{k_n=k_{min}}^{2\pi/\Delta_x} a_n\cos (k_nx - \omega_n t + \theta_n)$ where $\theta_n$ is the phase shift, assigned randomly to each mode.

\begin{figure}[h]
    \begin{center}
    \noindent\includegraphics[width=0.9\textwidth]{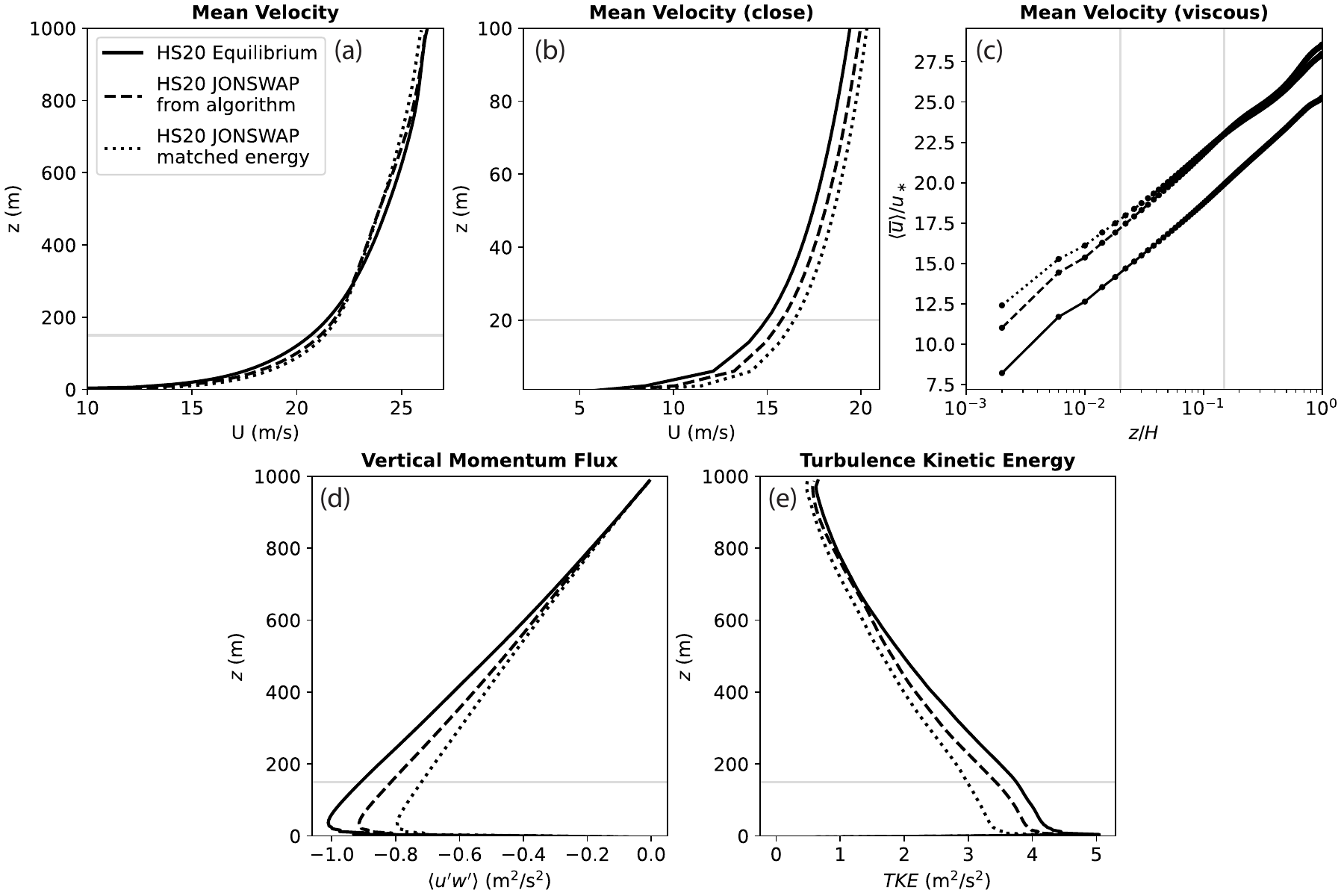}
    \caption{Mean velocity (a-c), vertical momentum flux (d), and turbulence kinetic energy (e) profiles for the high $H_s$ wave conditions described in Table \ref{tab:les_cases} and shown in Figure \ref{fig:les_spectra}. All cases shown in this figure were run with bulk velocity $U_{bulk} = 24$ m/s leading to a velocity at 10 m $U_{10}$ from 13.1 to 14.9 m/s. Lines to mark 20 m and 150 m above the sea surface are included for reference. 150 m is the turbine hub height for the IEA 15MW Reference Turbine \cite{iea_wind_definition_2020} and is used as an example of the impact different wave spectrum parameterizations would have on predictions for an example offshore wind turbine.}
    \label{fig:les_profiles}
    \end{center}
\end{figure}

The impact of the unresolved waves---those that are smaller than the LES grid size---is represented through a subfilter wave drag model. Subfilter drag is calculated with an equilibrium rough wall model \cite{anderson_affects_2020},
\begin{equation*}
    \widetilde{\tau}\big|_{wall} = \left[\frac{\kappa \left(\widetilde{u}_b-\widetilde{u}_s\right)}{\log\left(\frac{\Delta_z/2}{z_{0,\Delta}}\right)}\right]^2 \text{ ,}
\end{equation*}
where $\widetilde{u}_s$ is the surface wave orbital velocity, $\kappa$ is the von Karman constant, and $\Delta_z/2$ is the center of the bottom-most grid point. The wall roughness is $z_{0,\Delta} = \left[z_{0,s}^2 + \left(\alpha_w\frac{\widetilde{u}_b-c_{sf}}{\widetilde{u}_b+c_{sf}}\sigma_\eta^\Delta\right)^2\right]$, where $z_{0,s}$ is the smooth wall roughness (small magnitude), $c_{sf}$ is a characteristic subfilter wave speed, determined through the dispersion relation from a characteristic subfilter wavenumber based on the weighted energy in the subfilter region of the wave spectrum
\begin{equation*}
    k_{sf} = \frac{\int_{2\pi/\Delta_x}^{2\pi/\lambda_c} k^n\phi(k)dk}{\int_{2\pi/\Delta_x}^{2\pi/\lambda_c} k^{n-1}\phi(k)dk} \text{ ,}
\end{equation*}
where the capillary wavelength $\lambda_c$ defines the smallest waves considered. $\sigma_\eta^\Delta = \left(\widetilde{\eta^2} - \widetilde{\eta}^2\right)^{1/2}$ is the r.m.s. of the wave height fluctuation. $\alpha_w$ is a dynamic coefficient, calculated by requiring that the total drag (resolved plus subfilter) is grid-independent and equal at the LES (grid) filter scale and a test filter scale, typically twice the grid filter. More information about the wave model can be found in Aiyer et al. (2024) \cite{aiyer_dynamic_2024} and Williams et al. (2026) \cite{williams_sea_2026}.

We generate full-scale MABLs in a streamwise- and spanwise-periodic domain with dimensions $L_x \times L_y \times L_z = 3000\text{ m} \times 1260\text{ m} \times 1000\text{ m}$. The grid is uniform in the horizontal directions with cell dimensions $\Delta_x \times \Delta_y = 8\text{ m} \times 9.8\text{ m}$. We selected this streamwise resolution because it is able to resolve the relevant wave spectra while balancing computational expense. The vertical grid starts at the bottom of the domain with $\Delta_z = 4$ m, stays uniform until a height of $200$ m, after which it is stretched linearly by 3\% until the maximum cell size of $\Delta_{z,max}=15$ m is reached (at around 550 m) and maintained until the top of the domain. The flow is initialized uniformly at a bulk velocity $U_{bulk}$ with perturbations added to the bottom 20\% of the domain to facilitate the transition to turbulence. We ran three sets of wave cases at $U_{bulk} = 12, \:18,$ and $24$ m/s. The flow is driven with a constant mass condition that maintains the bulk velocity and allows the surface stress dynamics to evolve without prescription. The same $U_{bulk}$ results in different values of $U_{10}$ and $u_\ast$ depending on the effective drag imposed by the selected wave field through the wave spectra, as is discussed in Section \ref{sec:les}\ref{subsec:les_results}. $U_{bulk}$ therefore does not directly drive $U_{10}$ but is linked through the choice of spectrum. The bottom boundary is a wall with imposed drag to represent the ocean waves (described above) and the top of the domain has a free-slip boundary condition. We allow the MABL to develop until statistically stationary, around time $t \approx 12H/u_\ast$, where $H = L_z = 1000$ m is the eventual boundary layer height, and then run for another 10,000 s to collect statistics for analysis. We selected these particular MABL and wave conditions to be realistic to the Atlantic Coast of the United States near New Jersey and New York \cite{the_wavewatch_iii_development_group_ww3dg_user_2019, zhou_sea_2023}, a location where offshore wind farms have been proposed, to understand the potential impact of wave spectrum parameterization selection on simulations that may be used to predict farm power production.

\begin{table}
    \scriptsize
    \caption{Mean wind velocity and turbulence kinetic energy near the surface at 20 m followed by the same quantities at 150 m, which is the hub height for the IEA 15MW reference turbine, used an example offshore wind turbine. Percent differences in predicted velocity, power, and TKE based on wave spectrum parameterization are also noted here.}
    \begin{center}
    \begin{tabular}{| c |*{3}{c} |}
    \hline\hline
    & Equilibrium & JS, algorithm & JS, matched \\
    \hline
    $\langle u \rangle$ at 20 m (m/s) & 15.2 & 16.0 &  16.7 \\
    velocity percent difference at 20 m& - & 5.3\% &  9.9\% \\
    \rule{0pt}{5ex}    
    $TKE$ at 20 m (m$^2$/s$^2$) & 4.29 & 3.97 & 3.43 \\
    $TKE$ percent difference at 20 m & - & 7.5\% &  20.0\% \\
    \rule{0pt}{5ex}
    $\langle u \rangle$ at 150 m (m/s) & 20.66 & 21.13 &  21.33 \\
    velocity percent difference at 150 m& - & 2.3\% &  3.2\% \\
    power percent difference at 150 m & - & 7.0\% &  10.0\% \\
    \rule{0pt}{5ex}    
    $TKE$ at 150 m (m$^2$/s$^2$) & 3.76 & 3.44 & 2.99 \\
    $TKE$ percent difference at 150 m & - & 8.5\% &  20.4\% \\
    \hline
    \end{tabular}
    \end{center}
    \label{tab:les_diffs}
\end{table}

\subsection{LES Results}
\label{subsec:les_results}
Mean profiles of velocity $\langle\widetilde{u}\rangle(z)$ (with dimensional and viscous scaling), vertical momentum flux $\langle u'w'\rangle(z)$, and turbulence kinetic energy $TKE = \frac{1}{2}\left(\langle u'^2\rangle + \langle v'^2\rangle + \langle w'^2\rangle\right)$ are shown in Figure \ref{fig:les_profiles} for the high $H_s$ wave condition outlined in Table \ref{tab:les_cases} and bulk velocity $U_{bulk} = 24$ m/s. Mean velocity, vertical momentum flux, and TKE profiles follow the expected shape of a neutral atmospheric boundary layer \cite{stull_introduction_1988}. There is a visible difference between the mean velocities predicted with each spectrum parameterization, varying by 1.8 m/s at 10 m above the surface and 1.5 m/s at 20 m. Even  higher in the boundary layer at 150 m above the sea surface---almost $2\lambda_p$ above the waves and the hub height of the IEA 15MW Reference Turbine \cite{iea_wind_definition_2020}---there is a 0.67 m/s difference in the streamwise velocity. While this difference might seem modest, in the context of an offshore wind farm, the power output scales with velocity cubed and so could result in more drastically different estimates. In this example case, wave spectrum differences are associated with a 7-10\% change in predicted available power at hub height for the high $H_s$ case initiated with $U_{bulk} = 24$ m/s. Aggregated over an entire offshore wind farm, this would amount to a substantial difference in predicted available energy. The turbulence kinetic energy at hub height also changes between 8.5-20\% based on wave spectrum parameterization. This is significant because increased turbulence contributes to more frequent maintenance requirements on a turbine due to increased loads on the blades \cite{eggers_wind_2003}. Values of mean streamwise velocity and turbulence kinetic energy measured at 20 and 150 m above the sea surface are summarized in Table \ref{tab:les_diffs}.

\begin{figure}[ht]
    \begin{center}
    \noindent\includegraphics[width=0.7\textwidth]{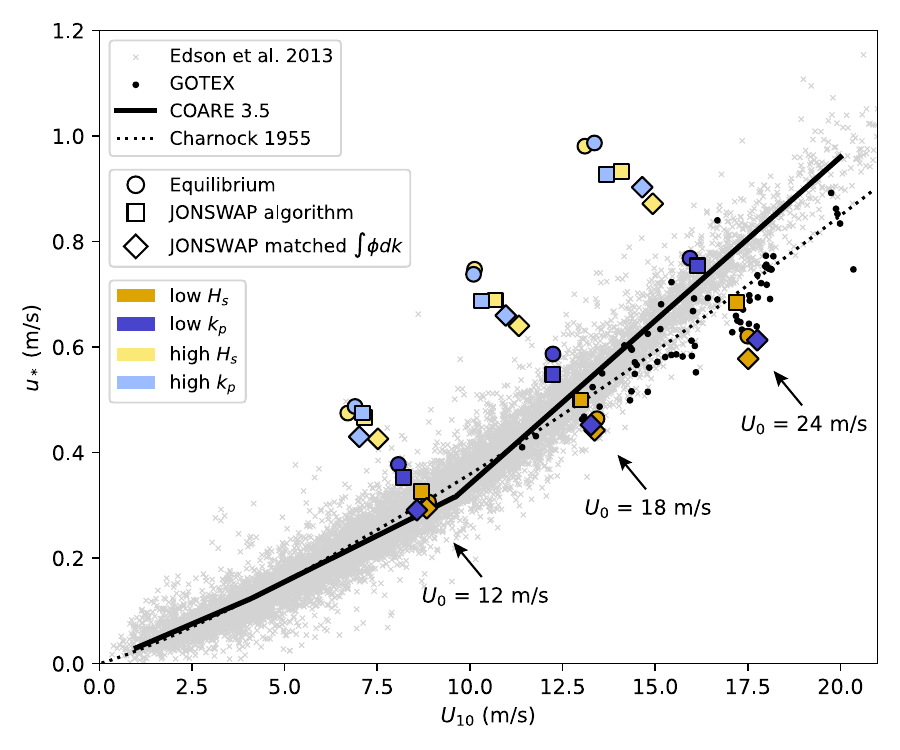}
    \caption{Surface roughness characterization through $u_\ast$ as a funciton of $U_{10}$ for the two wave conditions with different spectrum parameterizations compared to the COARE 3.5 fit and accompanying field data collated by Edson et al. (2013) \cite{edson_exchange_2013} as well as data from the GOTEX campaign \cite{romero_airborne_2010}. Movement across the COARE 3.5 line is consistent with changing wave conditions \cite{williams_sea_2026}}
    \label{fig:les_coare}
    \end{center}
\end{figure}

\begin{figure}[ht]
    \begin{center}
    \noindent\includegraphics[width=0.95\textwidth]{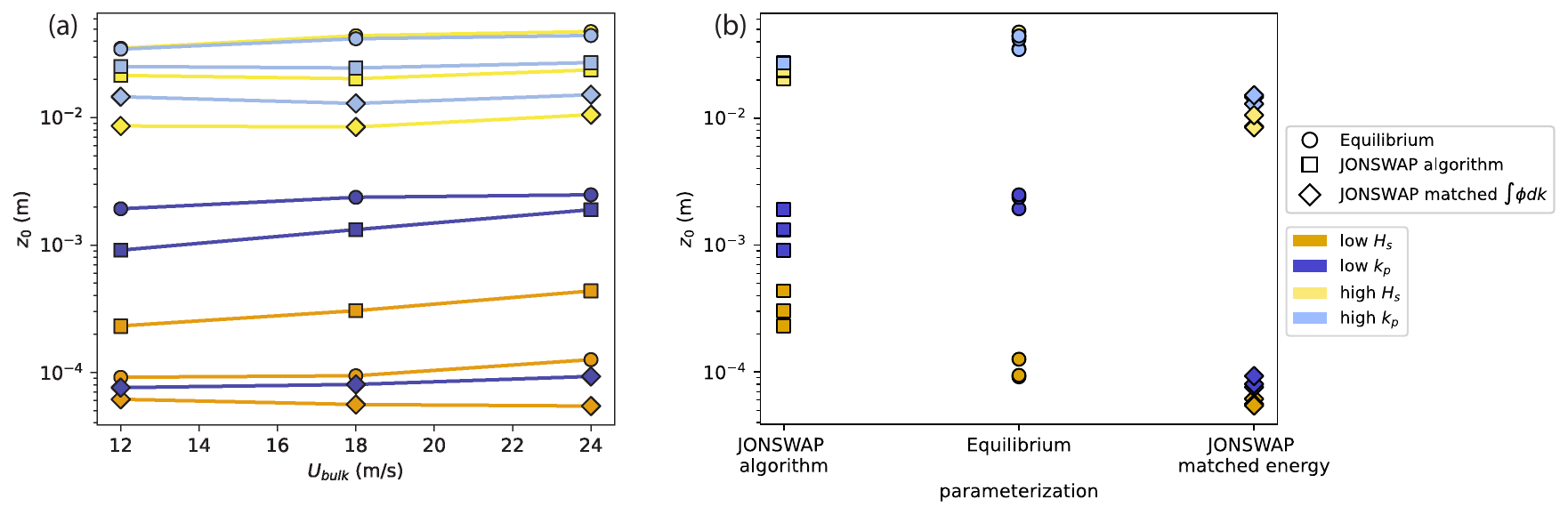}
    \caption{Equivalent global roughness values calculated from $u_\ast$ and $U_{10}$ through the logarithmic boundary layer relationship as a function of $U_{bulk}$ (a) and spectrum parameterization (b). Equivalent $z_0$ does not change much with bulk velocity because the wind and waves in this LES are one-way coupled; where it does, a slight increase in roughness can be attributed to the increase in slip velocity between the wind and waves. Equivalent roughness does change based on parameterization, in some cases by up to an order of magnitude. Wave conditions change whether the algorithm or matching energy provides a JONSWAP spectrum that is more similar to the Equilibrium one.}
    \label{fig:les_z0}
    \end{center}
\end{figure}

Furthermore, the drag characteristics of the resulting MABLs and their behavior close to the sea surface show a starker difference between spectrum parameterizations. Figure \ref{fig:les_coare} shows a plot of the friction velocity $u_\ast$ plotted as a function of $U_{10}$ from each of the simulations alongside field data collated by Edson et al. (2013) \cite{edson_exchange_2013} and those taken in the GOTEX campaign \cite{romero_airborne_2010}. We also include the relationship between $u_\ast$ and $U_{10}$ outlined by the COARE 3.5 fit for momentum exchange over the open ocean presented by Edson et al. (2013) \cite{edson_exchange_2013}. The lowest wave case (low $H_s$) results in very little variation in $U_{10}$, particularly when the wind is low, but a change in $u_\ast$ up to 15\% for $U_{bulk}$ = 24 m/s. For higher wave conditions, changes in $U_{10}$ become more substantial, resulting in a 13.6\% difference (13.1 m/s to 14.9 m/s) for the highest wind case. As the wind velocity increases, the sensitivity to the spectrum parameterization increases as well; the spread that results from changing the wave spectrum is akin to changing the wave conditions themselves, whereas increasing the wind speed results in movement parallel to the COARE 3.5 fit. A more thorough discussion about the impact of changing wind and wave parameters on the MABL can be found in \cite{williams_sea_2026}.

Assuming that the MABL takes shape according to the logarithmic relationship $U(z) = \frac{u_\ast}{\kappa}\log\left(\frac{z}{z_0}\right)$, we can compute an equivalent global surface roughness from the $u_\ast$ and $U_{10}$ shown in Figure \ref{fig:les_coare}. Figure \ref{fig:les_z0} shows the equivalent global $z_0$ plotted as a function of $U_{bulk}$ (\ref{fig:les_z0}a) and spectrum parameterization (\ref{fig:les_z0}b). There is minimal variation in $z_0$ with increases in bulk wind velocity; although this counters the canonical intuition that wave $z_0$ changes with wind velocity, it is important to note that this assumes wind-wave equilibrium (that is, that the waves are generated directly by the wind above them). We present an LES study in which the waves are set independently of the wind. Because ocean conditions are frequently in wind-wave disequilibrium because of sudden shifts in wind strength and direction \cite{sullivan_large-eddy_2008, edson_exchange_2013}, understanding the changes in roughness through this one-way coupled system is valuable. There is, in some cases, a slight increase in the equivalent $z_0$ with increased wind speed---this demonstrates that the waves also cannot be treated as static roughness elements through an equivalent $z_0$ because the relative velocity between wind and waves plays a role in the formulation of drag. The equivalent roughness changes by up to an order of magnitude based on the spectrum parameterization and definition. The sensitivity is higher for the cases with lower steepness $k_pH_s$ (low $H_s$ and low $k_p$ cases). Calculating the equivalent JONSWAP spectrum using the formulae provided by Hasselmann et al. (1973) \cite{hasselmann_measurements_1973} produces a roughness more similar to the Equilibrium spectrum for the high $k_p$, low $k_p$, and low $H_s$ cases (although the degree to which this is true varies) but matching the total energy results in a more similar roughness for the high $H_s$ case, making the it difficult to universally assess what the best method for calculating equivalency is. Overall, it is clear that the MABL global equivalent roughness is sensitive to the selection of spectrum parameterization.

\section{Discussion and Conclusion}
We compare data collected from four different field experiments in fetch-limited, fully-developed, and mixed ocean conditions to three ocean wave spectrum parameterizations. None of the three examined spectrum parameterizations are able to consistently capture the wave energy spectrum in mixed conditions very close to the shore because they were not developed to account for waves coming from multiple directions or in shallow areas. The Pierson-Moskowitz parameterization \cite{pierson_proposed_1964}, which depends on one input parameter $U_{19.5}$ and other fitted constants, consistently overestimates the energy in the wave spectrum, particularly near the peak and more dramatically so in fetch-limited wave conditions. The JONSWAP parameterization, presented by Hasselmann et al. (1973) \cite{hasselmann_measurements_1973}, captures the energy near the peak wavenumber in fetch-limited scenarios where wind-wave growth is defined by distance from a steady, identifiable wind source and fetch is small; JONSWAP concentrates more energy at the peak wavenumber for fully-developed sea conditions and falls off faster for high wavenumbers. The Equilibrium spectrum, based on theory from Toba (1973) \cite{toba_local_1973} and Phillips (1985) \cite{phillips_spectral_1985} and presented in this paper, does not capture the peak for fetch-limited waves (particularly those with small fetch) but improves as the conditions approach a fully-developed sea. In the fully-developed sea cases, the Equilibrium spectrum captures the location, magnitude, and shape of the peak as well as the high wavenumber fall-off better than the Pierson-Moskowitz and JONSWAP parameterizations.

We then demonstrate the importance of carefully selecting a wave spectrum parameterization by assessing the sensitivity of an MABL to the wave spectrum through an LES study. The LES data show that changing the wave spectrum parameterization can result in distinct changes in equivalent global roughness of the MABL, as well as measurable differences in the wind speed and turbulence kinetic energy both near the surface (10-20 m) and higher up in the boundary layer where, for example, and offshore wind turbine might be (150 m). In the context of many engineering studies (including the LES section in this paper), the goal is not to predict wave parameters based on the wind but instead to impose an accurate wave spectrum shape based on set or known wave parameters and investigate their importance on a specific engineering design. The LES study shows that the choice of spectrum parameterization makes a significant difference in this context.

As conclusions, or recommendations, we note that i) JONSWAP should not be used systematically as it only describes limited data sets; ii) there is a sensitivity to the choice of spectrum in various applications, here illustrated by the impact on the MABL; and iii) while the Equilibrium spectrum is best-suited for a fully-developed sea, there is no (simple to use) universal parameterization that works for every type of sea-state. Therefore, careful selection of a spectrum parameterization is vital especially in cases that have to do with large-scale prediction of air-sea interactions. A quantitative evaluation of the different possible parameterizations would require the systematic analysis of a much larger data set.

In addition to the changes in accuracy given different wave conditions between the three spectrum parameterizations, data for $k_p$ and $H_s$ are more readily accessible (while fetch is more difficult to assess in the open ocean), so reliably tailoring wave spectra to specific sites is easier using the Equilibrium spectrum parameterization. Although the JONSWAP spectrum uses two input parameters, it still relies on the three-way coupling between fetch, wind velocity, and wave conditions (equations \ref{eq:js_omega}-\ref{eq:js_int}) that constrains the wavelength and energy, which fails in most practical conditions. Using inputs that directly control wave parameters allows for more flexible and precise definition of the wave spectrum without confusion about whether the waves are in equilibrium with the wind. Challenges in representing the wave spectra simply for a wide range of conditions have motivated extensive operational (spectral) modeling development \cite{romero_distribution_2019, the_wavewatch_iii_development_group_ww3dg_user_2019, ecmwf_ifs_2019, zhou_sea_2023,sauvage_improving_2023, sauvage_fetch-dependent_2025, ardhuin_sizing_2025, liu_swell_2026}, which are then extensively evaluated for the peak of the wave spectrum and significant wave height at given wind speed. An accurate spectral parameterization can be used based on simple products or algorithms to predict a wave spectrum's main characteristics $H_s$ and $k_p$ \cite{wang_scattering_2025, hell_particle--cell_2025}.

Scientific inquiry and engineering design both employ spectrum parameterizations; data from the field and LES study suggest that the choice of wave energy spectrum should be a careful one based on the geography of a specific site (fetch-limited versus fully-developed seas) and precise wave conditions.

\section*{Acknowledgments}
The authors gratefully acknowledge financial support from the Princeton University Andlinger Center for Energy and the Environment, High Meadows Environmental Institute, and the New Jersey Wind Institute Fellowship Program. This work is supported by the National Science Foundation under grant 2318816 to LD (Physical Oceanography program).

%
%
\section*{Data Statement}
The data that support the findings of this study are available at https://github.com/hhwilliams/ocean-wave-spectrum-parameterizations. 

\clearpage
\bibliography{references}

\end{document}